# Database of semiconductor point-defect properties for applications in quantum technologies

Vsevolod Ivanov[1,2,†], Alexander Ivanov[3], Jacopo Simoni[1], Prabin Parajuli[1], Boubacar Kanté[4,5], Thomas Schenkel[2], Liang Tan[1]

[1]Molecular Foundry, Lawrence Berkeley National Laboratory, Berkeley, California 94720, USA

[2]Accelerator Technology and Applied Physics Division, Lawrence Berkeley National Laboratory, Berkeley, California 94720, USA

[3]Department of Computer Science, Brown University, Providence, Rhode Island 02906, USA

[4]Department of Electrical Engineering and Computer Sciences, University of California Berkeley, Berkeley, California 94720, USA

[5]Materials Sciences Division, Lawrence Berkeley National Laboratory, Berkeley, California 94720, USA

† vivanov@lbl.gov

**Abstract**

Solid-state point defects are attracting increasing attention in the field of quantum information science, because their localized states can act as a spin-photon interface in devices that store and transfer quantum information, which have been used for applications in quantum computing, sensing, and networking. In this work we have performed high-throughput calculations of over 50,000 point defects in various semiconductors including diamond, silicon carbide, and silicon. Focusing on quantum applications, we characterize the relevant optical and electronic properties of these defects, including formation energies, spin characteristics, transition dipole moments, zero-phonon lines. We find 2331 composite defects which are stable in intrinsic silicon, which are then filtered to identify many new optically bright telecom spin qubit candidates and single-photon sources. All computed results and relaxed defect structures are made publicly available online at quantumdefects.com, a living database of defect characteristics which will be continually expanded with new defects and properties, and will enable researchers to select defects tailored to their applications.

## Introduction

Solid-state point defects are atomic-scale imperfections that can occur naturally or artificially in materials, leading to unpaired electrons or nuclear spins.  Against the background of the host structure, isolated defects behave analogously to small molecules, with associated quantum optical and magnetic properties that are increasingly being harnessed for applications in quantum information science (QIS), including quantum computing [1,2], networking [3], navigation [4], and sensing [5]. In particular, color-center defects in silicon have several advantages for QIS applications, including emission of photons in the telecommunication range, long electron spin coherence times, and narrow linewidths. Furthermore, the prevalence of silicon-based telecommunications and computing devices means that quantum devices based on silicon color-centers could readily be integrated with existing technologies and take advantage of well-established manufacturing processes for silicon electronics and photonics.

Several defects have received considerable attention due to their ease of synthesis, spin states, or optical properties including the nitrogen-vacancy center in diamond, divacancy in silicon carbide, and G-center in silicon. There are a number of excellent reviews that summarize the experimental and theoretical progress on known solid state point defects [6-10]. There have also been several attempts to systematically study defects in materials, though these have been limited in scope, focusing either on a limited number, or specific type of defect [11-14]. In fact, many studies focus exclusively on defects composed of single elements or vacancies even though many known defects are composite, involving as many as four interstitial or substitutional sites [10]. Despite these efforts, known color-center defects don't have optimal properties for QIS applications, and adequate candidate defects are yet to be identified [6].

First-principles density functional theory methods are an essential tool in the study of quantum point defects, allowing experimental results to be interpreted in terms of the computed structural, electronic, and optical properties of a defect. Numerous methods for computing the observable properties of defects have been developed [15-18]. When combined with high-throughput calculations, these methods can become predictive, guiding the discovery of new defect candidates, or give general insights about the physical and chemical trends in defect properties.

In this paper we perform high-throughput calculations of defects in several different hosts, including diamond, silicon carbide, and silicon. Focusing on defects in silicon, we combinatorically generate over 10,000 defect structures, including composite defects, and identify several candidate defects with optimized properties for their application as spin-qubits or as single-photon sources. The procedure is performed in several stages outlined in Fig. 1, with each stage having increasing computational cost. The results are organized into a database of defect properties, publicly available online at quantumdefects.com. We briefly describe the stages below, and provide more detailed explanations and results in the Supplementary Information.

**Computational Approach**

Generally, two different approaches can be taken for identifying new kinds of defects - modifying known defects, or discovering all-together new defects from permutations of elements and vacancies. The latter category of defects can be experimentally synthesized using a combination of ion beams to precisely implant atoms [19,20], and annealing or irradiation to add or remove additional damage to the crystal lattice. The search space of such possible defects is extremely large, since there are defects that contain up to four atoms [10]. In the first stage of our high-throughput search, we combinatorically generate all possible configurations composed of between 1 and 4 atom clusters placed at substitutional and interstitial sites, up to a total of four atoms contained within a single unit cell. Configurations that are equivalent by symmetry are eliminated by finding those that have equivalent sites and equivalent distances between sites. For silicon, this results in 493 distinct configurations, including 2 one-atom, 11 two-atom, 73 three-atom, and 407 four-atom configurations. Additional details on how configurations are generated can be found in Supplementary Information appendix A.

Selection of candidate defect structures is informed by practical results from experiment. The creation of defects is done by introducing elements into the host lattice either by diffusion or ion beam implantation. We consider 50 elements (more details in the Supplementary Information), which were used to populate the previously described configurations to create initial defect structures. For all host materials, we consider defects comprised of a single element at an interstitial or substitutional site. For silicon, a more thorough search of composite defects is performed. This includes defects where these two types of sites are "dressed" with nearby atoms native to natural silicon, namely carbon, oxygen, hydrogen, silicon, as well as vacancies. These use the 11 distinct two-atom configurations that were previously determined combinatorically, leading a total of 2600 structures. Additionally, three configurations – two adjacent interstitials, two adjacent substitutionals, and an adjacent interstitial and substitutional – are used to generate all combinatorically allowed defects containing two of the selected elements, yielding 5050 structures. Finally, we perform a depth-first search of potential defects by considering all one-four atom configurations comprised of a single element for select 2nd row elements, as well as germanium, phosphorus, and halides, for an additional 3944 structures. More details can be found in Supplementary Information appendix B.

In the second stage of the procedure, the formation energy of each defect is computed for different positions of the Fermi level, to determine the range of stable charge states and the most stable charge state of each defect in the intrinsic semiconductor. Each defect is embedded into a 3 x 3 x 3 supercell of the host material, and the electronic structure is computing using VASP [21,22] with PBE functionals [23]. Calculations are performed at the Γ-point with an energy cutoff of 300 eV, converging to a tolerance of $10^{-8}$ eV, and the structure is relaxed to a force tolerance of 0.01 eV/Å. Initially, the charge states -2, -1, 0, +1, and +2 are used, and after formation energies are computed along with stability range within the gap, additional charge states are computed until all stable charge states within the semiconductor gap are found. Formation energies are computed using

$$\Delta E^f[D, q] = G[D, q] - G[\text{bulk}] - \sum_i n_i \mu_i + q(E_F + \epsilon_{VBM}) + E_{corr}$$

(1)

where $\Delta E^f$ is the formation energy for defect $D$ in charge state $q$, $G[D, q]$ and $G[\text{bulk}]$ are the free energies of the defect-containing and pristine supercells respectively, $n_i$ is the number of atoms with chemical potential $\mu_i$ added or removed to create the defect, $\epsilon_{VBM}$ is the energy eigenvalue of the highest occupied band, $E_F$ is the Fermi level which can take any value within the material band gap, and the $E_{corr}$ corrects errors due to the finite size of the supercell. This formula is implemented in the Spinney package [24], with Kumagai and Oba [25] scheme for the $E_{corr}$ term. Further details on this stage are given in Supplementary Information appendix C.

In the third stage, the structures of defect charge states which are stable within the gap are used as the starting point for a ground state structural relaxation using the Heyd–Scuseria–Ernzerhof hybrid functional (HSE06) [26]. This relaxation is performed at the Γ-point with an energy cutoff increased to 400 eV and the energy and force thresholds decreased to $10^{-10}$ eV and 0.001 eV/Å respectively. The resulting ground state atomic configuration $q_{GS}$ is subsequently used for a final calculation which computes the complex wavefunctions, density of states (DOS), zero-field splitting (ZFS), and hyperfine couplings. Real-space wavefunctions were extracted using the VASPKIT package [27], which we then used to compute the spread of the wavefunction to determine the localization of each state, and then draw energy-level diagrams describing the electronic structure of each defect. Based on the position of the localized defect levels, we classified the defects as "Buried" if there were no defect levels within the bulk gap, "Valence Band Buried" if the lowest unoccupied defect level was above the conduction band minimum, "Conduction Band Buried" if the highest occupied defect level was below the valence band maximum, and "Midgap" if both the highest occupied and lowest unoccupied defect levels lay within the bulk gap. Converged ground state configurations $q_{GS}$, electronic structures, and properties at the HSE06 level of computation were obtained for 4291 defects. These were selected based on their stability in undoped silicon and having a small charge state ($|c| < 2$), as heavily charged defects will exhibit strong electrostatic interactions and would require large energy corrections for various computed parameters due to finite size effect [18,24,25]. Details on the third stage and postprocessing are given in Supplementary Information appendix D.

The fourth stage computes the excited state properties for each defect using the constrained occupation method [28,29] at the level of HSE06 and the same computational parameters used for the third stage. The zero-phonon line (ZPL) is computed as the total free energy difference between the ground and excited states, and the relaxed atomic configuration of the excited state, $q_{EX}$, is used for postprocessing to extract other relevant excited state quantities. For each defect, the relaxation procedure and postprocessing is repeated for different excited states, in order of priority: lowest energy spin-allowed transitions between localized defect states, lowest energy spin-allowed transitions between a localized defect state and bulk state, higher energy local-local or local-bulk transitions, and finally spin-forbidden transitions. The constrained occupation method does not always yield a converged excited state energy because of inherent issues in ordering energy eigenvalues [29]. Of the defects for which ground state calculations were completed in the previous stage, we complete the fourth stage for 2331 identified defects which have excited states for which the constrained occupation method has stable convergence. It should also be noted that ZPLs computed in this way may have an error on the order of 100 meV due to several factors, such as finite-size effects due to the periodicity of the supercell or using a Γ-point only calculation, which can be reduced by increasing the supercell size or using a larger **k**-point grid [29]. When considering transition to delocalized bulk states, there are also corrections on the order of 100 meV due to finite-size effects [18], which are not considered here. Supplementary Information appendix E gives further details on this stage.

In the final stage, defect phonon properties [30] and displacement gradients are computed and used in the calculation of photoluminescence spectra [29,31], linewidths, and dephasing times [32]. The computational demand of this stage significantly exceeds that of previous stages, and so is only done for defects which are optically bright, have non-trivial spin ground states with large zero field splitting, or have zero phonon lines near or in the telecom band. Further computational details may be found in Supplementary Information appendix F, but the analysis of phonon-related properties, including linewidth, photoluminescence spectra, and dephasing times is beyond the current scope, and will be covered in a later work.

All computed defect properties are made publicly available through a searchable web interface hosted at quantumdefects.com. Each defect charge state has its own distinct data page, which also contains plots of the energy levels and chemical potential-dependent formation energies, as well as visualizations of relaxed ground and excited state configurations and localized wavefunctions when available. Further details on the defect properties derived in stages 2-5 are given in Supplementary Information appendix G.

**Results and Discussion**

Our methodology enables us to rapidly screen for defects tailored to specific applications. For instance defects that act as telecom single photon sources need to be optically bright and necessarily have a ZPL emission within the telecom band (763 meV - 984meV). Spin qubit candidates should additionally possess a spin-degree of freedom, which would allow for optically addressable spin states. In Fig. 2 we provide a small selection of telecom silicon defects identified by our database. Supplementary appendix H gives more details and provides tables of candidate defect targets for synthesis.

A more general overview of discovered defects is given in Fig. 3. Defects are classified by spin ground state, defect level type, ZPL, and TDM. Several high-quality defect candidates are highlighted that are optically brighter than the well-known silicon G-center and T-center [10, 29], have spin doublet or triplet ground states, telecom ZPL, and midgap defect levels. While it has been proposed that having both participating defect levels localized completely within the bulk gap is not a necessary requirement for QIS applications [33], the localization of both levels in a transition is nevertheless a crucial factor in determining defect properties, with delocalized states potentially leading to lower ZFS, large Huang-Rhys factors, and shorter dephasing times due to increased electron-phonon coupling [29].

## Conclusion

We have performed a high-throughput search encompassing over 50,000 defects in several common host materials, and identified a large number of optically bright defects with non-zero spin ground states and telecom zero-phonon-lines. These defects are suitable candidates for spin qubits or single-photon sources, which can be used for QIS applications. A statistical analysis of computed defects is used to identify which elements should be implanted when seeking defects with particular properties, establishing guidelines for experimental synthesis efforts. The computed properties and structures for all considered defects are made freely available through a web application hosted at quantumdefects.com.

The next critical steps in the study of point defects in materials involve determining which defects are most favorable to form and under what conditions. Formation energy data can be used to compare the relative stability of composite defects and single-atom defects, while formation rates can be estimated from molecular dynamics simulations. However, for applications in quantum information science (QIS), more stringent constraints on defect properties are required, such as indistinguishable photons and long dephasing times. This necessitates the computation of linewidths and dephasing times to screen optimal defect candidates. The results of these calculations will need to be carefully corroborated with experimental results to ensure the predictive capacity of these methods.

Another crucial next step is to use a more suitable theory, such as Dynamical Mean Field Theory to study lanthanide defects. In these defects, the strong coupling between nuclear and electronic degrees of freedom allows them to be used as quantum memories. While most of the extensive work has been done on silicon, it is essential to explore other materials as well, considering their other potential applications besides QIS. The computed properties in the database can be used to identify point defects that can be used for chemical sensing of targeted molecules, understanding structural stability and radiation damage in materials, and identifying clean dopants in semiconductors, which when implanted will not create states within the gap. These potential applications demonstrate the utility of the defect property database beyond QIS, and will hopefully accelerate discovery across a wide range of problems related to defects.

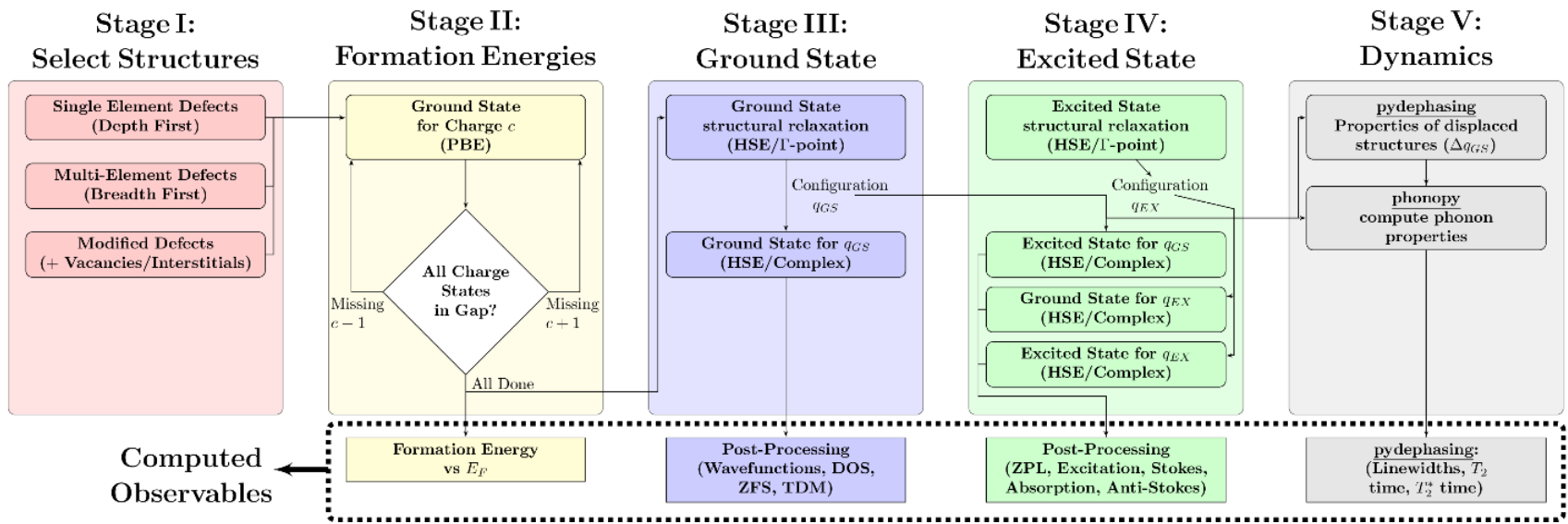

**Fig. 1. Workflow for computing defect properties.** The figure shows the various stages used in our high-throughput search of solid-state point defects. Postprocessing in stages II-V is used to obtain observable properties of defects, including DOS – density of states, ZFS – zero-field splitting, TDM – transition dipole moment, ZPL -zero-phonon line.

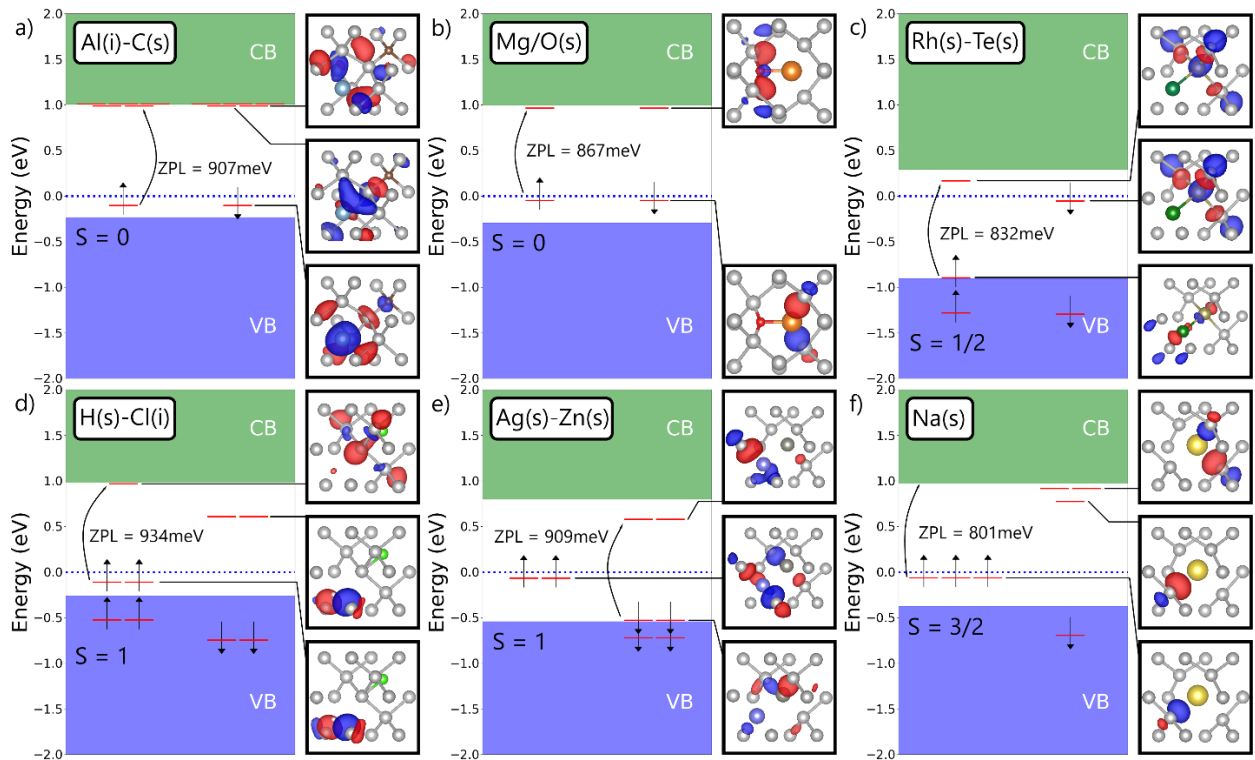

**Fig. 2. Several telecom defects identified by the high throughput search.** Panels a-f) show the computed electronic structures of several selected defects with telecom ZPL, few defect states within the gap, and either non-trivial spin ground state or midgap localized states. The conduction band (green), valence band (blue), Fermi level (dashed line), localized defect states (red lines), as well as spin up/down defect level occupations (left/right side) are indicated. To the right of each energy level diagram are shown the defect structures with superimposed visualizations of the wavefunctions corresponding to the marked defect levels. Defect names indicate which atoms compose the defect, sitting at substitutional (s) or interstitial (i) sites.

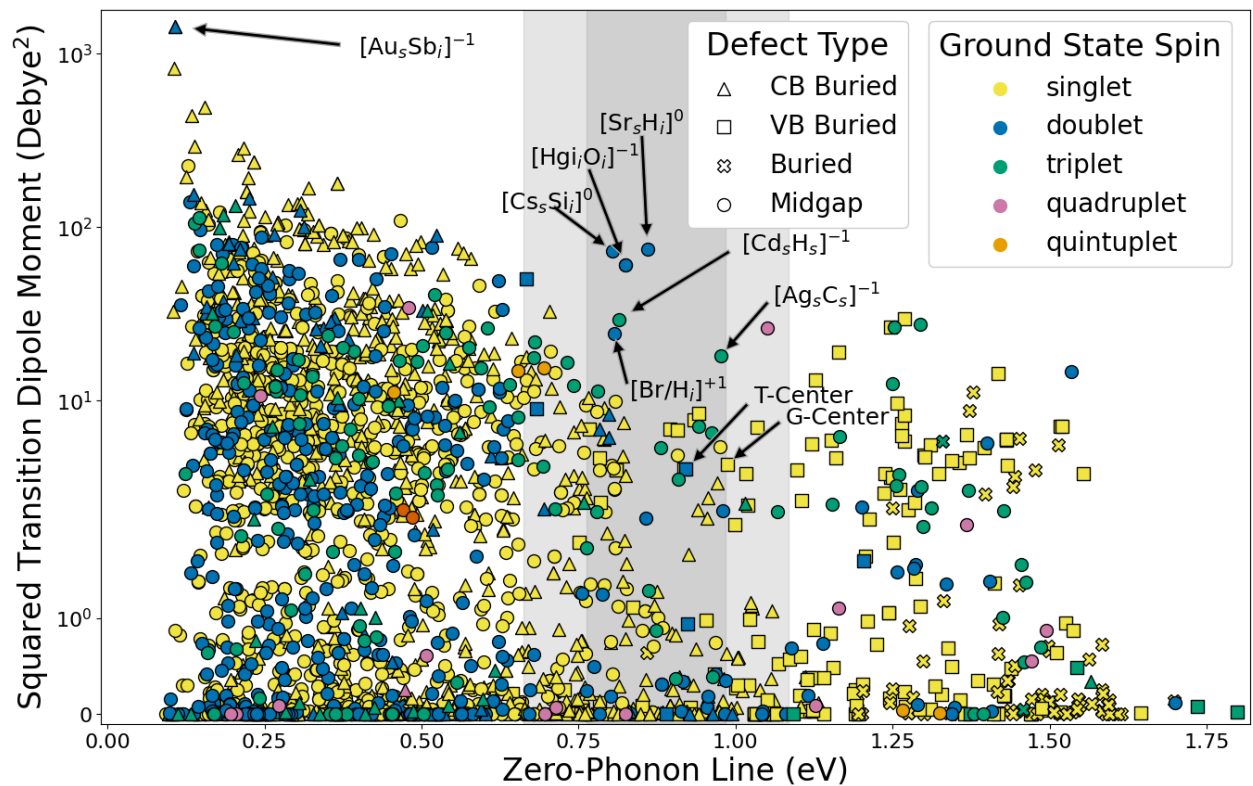

**Fig. 3. Distribution of silicon defects identified by the database.** The squared transition dipole moment of defects is plotted versus their zero-phonon-line. Marker shape indicates whether this excited state for the defect involves two midgap states (Midgap), a midgap state and a level within the conduction band (CB Buried), a midgap state and a level within the valence band (VB Buried), or no midgap levels (Buried). Marker color denotes the number of unpaired spins in the ground state – 0 (yellow), 1 (blue), 2 (green), 3 (magenta), or 4 (orange). The dark grey region marks the extent of the telecom band, while the light grey region indicates the extent of the 0.1 eV error potentially present due to the computation limitations. Specific high-quality defects, along with the T-center and G-center are indicated by arrows.